# Inverse Design of Chiral Structures for Giant Helical Dichroism

Chia-Chun Pan[1,*], Munseong Bae[2,*], Hongtao Wang[3], Jaesung Lim[2], Ranjith R Unnithan[1], Joel Yang[3], Haejun Chung[2,†], Sejeong Kim[1,†]

[1]Department of Electrical and Electronic Engineering, University of Melbourne, VIC, Australia
[2]Department of Electronic Engineering, Hanyang University, Seoul, Korea
[3]Engineering Product Development, Singapore University of Technology and Design, Singapore

*These authors are equally contributed.
Corresponding authors: haejun@hanyang.ac.kr, sejeong.kim@unimelb.edu.au

## Abstract

Investigating chiral light-matter interactions is essential for advancing applications in sensing, imaging, and pharmaceutical development. However, the chiroptical response in natural chiral molecules and subwavelength chiral structures is inherently weak, with the characterization tool limited to optical methods that utilize the light with spin angular momentum (SAM). To overcome this, orbital angular momentum (OAM) beams, characterized by helical wavefronts, have emerged as a compelling research focus. Helical dichroism (HD) describes the differential absorbance of OAM beams with opposite signs of topological charges. By using inverse design with adjoint methods for topology optimization, we design the chiral structure optimized to increase HD response under OAM beam incidence, demonstrating a giant HD response of ~107% with topological charges $|\pm\ell| = 3$ at the wavelength of 800 nm. This study reveals distinct helicity-dependent interactions between the structure and OAM beams, highlighting the potential for custom-tuned chiroptical responses.

## Introduction

Chirality refers to a geometric property of objects where they cannot be superposed onto their mirror images. Identifying chirality is important, particularly for chiral molecules, also known as enantiomers, as their opposite-handedness may have drastically different effects on biological systems [1-3]. Therefore, the study of the interactions between light and chiral materials has gained momentum as a method to differentiate one enantiomer from another, with the majority of research efforts focusing on light with spin angular momentum (SAM), i.e., circularly polarized light [4-6]. Chiral molecules preferentially absorb either left-handed circularly polarized (LCP) light or right-handed circularly polarized (RCP) light, resulting in an intensity contrast in the transmission or reflection spectra. This intensity contrast is known as circular dichroism (CD) [4, 7, 8]. However, the CD contrast is often weak when measured from natural chiral molecules due to the small size of the target molecules relative to the wavelength of incident light and the beam spot size [9-11]. Chiral metamaterials, which are artificially created, can address this issue and enhance the CD contrast. Many studies have reported the design and optimization of chiral nanostructures that exhibit a large CD contrast [12-14].

Recent findings suggest that utilizing light beams with orbital angular momentum (OAM) can offer greater chiral sensitivity compared to beams carrying SAM [15]. OAM possesses the unlimited degree of freedom defined by the topological charge denoted as $\ell$ ($\ell$ can be any integers) [16-18]. Similar to CD measurements, chiral molecules show different absorption for OAM beams of opposite handedness, resulting in an intensity contrast in the spectrum, a phenomenon referred to as helical dichroism (HD) [19, 20]. J. Ni et al., measured the HD of a chiral structure as a function of the varying topological charges of OAM beams, achieving an experimental HD of 20% [21], followed by a report by N. Dai et al., observing a further increase in the HD signal to 50% [22]. The infinite degree of freedom afforded by OAM beams makes them particularly advantageous for probing complex chiral systems, optimizing light-matter interactions, and designing advanced chiroptical devices, thus opening new avenues for exploring and engineering chiral phenomena.

The design of chiral nanostructures to enhance HD has largely followed similar structural designs used for CD, often emphasizing intuitively chiral geometries such as spiral or helical forms

[21-23]. However, no established design strategy specifically targeting the enhancement of HD currently exists, leaving the process largely dependent on intuition. Inverse design provides a robust framework for photonic design by formulating the process as an optimization problem to achieve the desired optical response. In contrast to traditional design methods, which typically depend on intuition, trial-and-error, or prior experience, inverse design enables the exploration of non-intuitive photonic design spaces with enhanced functionality, allowing for the realization of compact and high-performance devices. Specifically, topology optimization with adjoint sensitivity analysis, commonly referred to as adjoint optimization and facilitated by the Born approximation and Lorentz reciprocity [24, 25], has been successfully applied to multifunctional photonic devices exhibiting high functionalities [26-28]. This approach proves particularly effective for designing structures that enable precise control over subwavelength-scale light-matter interactions tailored to a specific functionality. By leveraging the potential of inverse design, custom-tuned chiroptical responses with high HD can be accurately realized.

In this work, we report the application of inverse design with adjoint methods to iteratively develop an HD chiral structure that maximizes the reflectance intensity contrast for positive and negative OAM topological charges. The induced HD responses for topological charges $|\ell| = \pm 3$ reached approximately 107%, a value that was below 50% in simulations with two-dimensional structures in previous works. This novel approach lays the groundwork for next-generation chiral structures, broadening their applications from chirality-sensitive devices and chiroptical spectroscopy to advancements in drug development.

## Results and Discussion

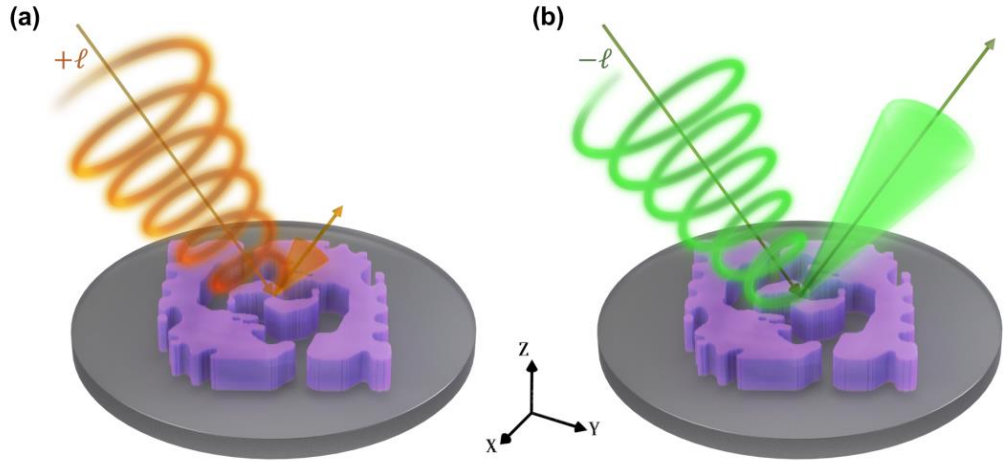

**Figure 1.** Schematic illustration of the chiroptical response between OAM beams and an inverse-designed chiral structure. **(a)** Interaction between a vortex beam with a positive topological charge (+ℓ) and the chiral structure (purple). The structure, made of $Si_3N_4$, which is placed on a glass substrate (grey plate). The arrows indicate the directions of the incident and reflected light. **(b)** Interaction between a vortex beam with a negative topological charge (-ℓ) and the chiral structure. The size contrast of the reflected beam response is proportional to the HD signal; specifically, the interaction with the positive topological charge exhibits stronger light absorption, resulting in a weaker reflected signal. The OAM beams are conducted at normal incidence.

Our proposed inverse design method optimizes the chiral structure to maximize the helical dichroic response. While many chiral structures have been designed with metals due to their ability to provide a strong chiroptic signal in a small footprint, dielectric structures can offer advantages as they are not subject to quenching near the surface. In this work, SiN is selected as the material for the chiral structure. It has a relatively high refractive index ($n \approx 2.0$) within its transparency window and shows nearly zero absorption across a broad range of wavelengths, including the visible. Additionally, SiN is a dielectric material with mature fabrication technology that is CMOS-compatible. The chiral structure preferentially interacts with one handedness over the other, leading to differences in reflectance intensity. **Figure 1a** conceptually illustrates that the inverse-designed structure interacts favorably with the OAM beam carrying a positive topological charge (depicted as orange helical light), leading to a lower reflected light intensity. Conversely, **Fig. 1b** shows the weaker interaction between a beam with a negative topological charge (represented as green helical light) and the same chiral structure shown in **Fig. 1a**. This weaker interaction results in less light absorption, which causes a smaller reduction in the intensity of the reflected light beam. HD is calculated from the reflectance difference between the positive ($R_{\ell+}$) and negative ($R_{\ell-}$) topological charges in relation to their mean values using the equation expressed as [21, 22]:

$$HD\ (\%) = \frac{R_{\ell-} - R_{\ell+}}{(R_{\ell-} + R_{\ell+})/2} \times 100\% \qquad (1)$$

This equation is used in other papers to define HD as well, with a value range of 0 to 200 %.

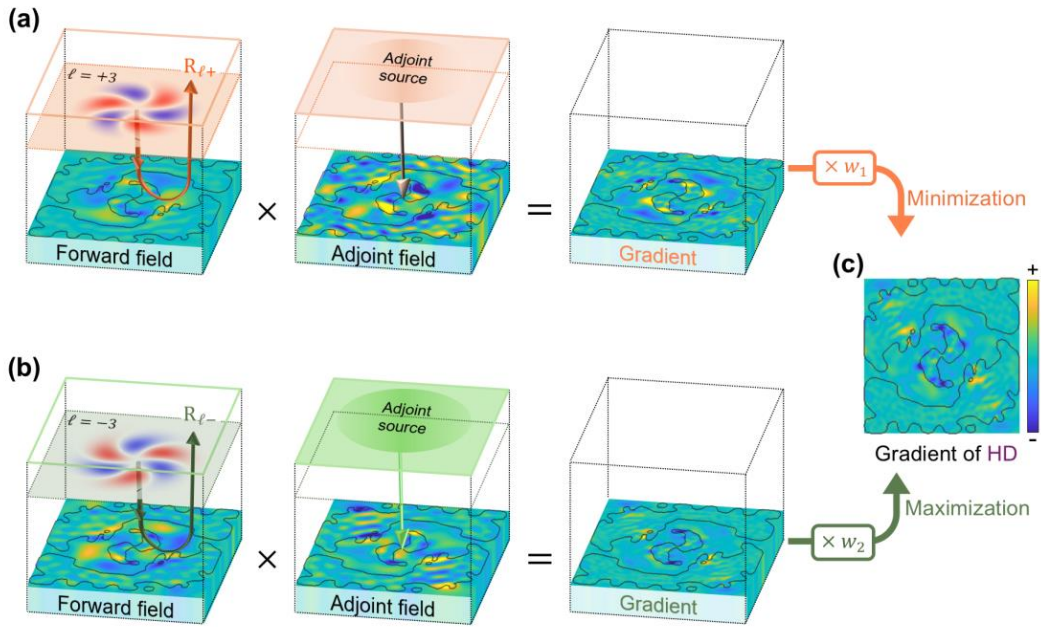

**Figure 2.** The adjoint optimization process within a single iteration cycle for maximizing helical dichroism (HD). In each iteration, sequential simulations compute the forward and adjoint fields, and their matrix product determines the reflectance gradient with respect to material density between clad and core materials. **(a)** The gradient of the reflectance ($R_{\ell+}$) for a helical incidence beam with a positive topological charge is weighted by a negative factor ($w_1$) to minimize $R_{\ell+}$. **(b)** The gradient of the reflectance ($R_{\ell-}$) for a helical incidence beam with a negative topological charge is weighted by a positive factor ($w_2$) to maximize $R_{\ell-}$. **(c)** The final HD gradient,

combining the two weighted gradients, directs the design process to enhance the difference between $R_{\ell+}$ and $R_{\ell-}$.

In this work, three-dimensional (3D) numerical simulations are conducted using the finite-difference time-domain (FDTD) method, implemented through MEEP, an open-source simulation tool [29]. **Figure. 2** illustrates the two subsets of adjoint gradient analysis, consisting of forward and adjoint simulations for positive $\ell$ (**Fig. 2a**) and negative $\ell$ (**Fig. 2b**), respectively. The adjoint gradient is computed as the matrix product of two electromagnetic field distributions within the design region: the forward field induced by the input OAM source and the adjoint field generated by the corresponding adjoint source. To simulate the OAM beam propagating along the z-direction in the simulation, Laguerre Gaussian (LG) mode beam is used where the equation is shown below [30]:

$$u_{\ell p}(r, \phi, z') = \frac{C}{w(z')} \left[\frac{r\sqrt{2}}{w(z')}\right]^{|\ell|} L_p^{|\ell|}\left(\frac{2r^2}{w^2(z')}\right) exp\left(\frac{-r^2}{w^2(z')}\right) exp(-i\ell\phi)$$

$$\times exp(-i\psi(z'))exp(ik\frac{r^2 z}{2(z'^2+z_R^2)})\hat{x} \quad (2)$$

where $C$ is a normalization factor: $\sqrt{\frac{2}{\pi(p+|\ell|)!}}$, $L_p^{|\ell|}$ is the associated Laguerre polynomials, $r = \sqrt{x^2 + y^2}$ is the radial distance, $w(z')$ is the beam width at $z'$, $\psi(z') = (|\ell|+1)\arctan(\frac{z'}{z_R})$ is the Guoy phase. Here, $z' = z_0 - z$ where $z_0$ is the position of the LG beam waist. Unlike circularly polarized light, which is inherently limited to two states, the topological charges of OAM beams, denoted by $\ell$, can theoretically span an infinite range of integer values. The infinite degrees of freedom offered by OAM beams make them particularly advantageous for probing complex chiral systems and designing advanced chiroptical devices. The wavelength of the incident OAM beam is set to $\lambda = 800$ nm, with the beam waist $w(z_0) = 800$ nm. The height of the SiN layer is set to 800 nm, with a minimum feature size of 200 nm, well above the smallest patterns that e-beam lithography can achieve. The SiN layer is sitting on a 600 nm-thick $SiO_2$ substrate with a refractive index of 1.45.

The figure of merit (FoM) for each subset is defined as the reflectance of the helical incidence beam with positive and negative topological charges ($R_{\ell\pm}$) as follows:

$$R_{\ell\pm} = \left|\frac{\int (E_{refl} \times H_{refl}^*) \cdot dS}{\int (E_{inc} \times H_{inc}^*) \cdot dS}\right|_{\pm\ell} \qquad (3)$$

where $\ell$ represents the topological charge of the helical incidence beam, $E_{refl}$ and $H_{refl}$ are the reflected electric and magnetic fields at the output monitor, $E_{inc}$ and $H_{inc}$ are the incident electric and magnetic fields at the output monitor for normalization, and $S$ denotes the size of the output monitor. The amplitude of the adjoint source is determined from the partial derivative of $R_{\ell\pm}$ with respect to the electric field within the output monitor, obtained from the forward simulation. **Figure 2c** shows the final gradient for HD maximization: a weighted summation of the two calculated gradients with positive and negative factors. The weight factors for the gradient of $R_{\ell\pm}$ in each iteration are determined as follows:

$$w_1 = \frac{-R_{\ell+}}{R_{\ell+}+R_{\ell-}}, \qquad w_2 = \frac{R_{\ell-}}{R_{\ell+}+R_{\ell-}} \qquad (4)$$

Where $w_1$ is the weight factor for $R_{\ell+}$ minimization, and $w_2$ is the weight factor for $R_{\ell-}$ maximization. The simulation volume for the inverse design is defined as 5 μm × 5 μm × 2.3 μm, encompassing the SiO$_2$ substrate and the OAM sources with targeted topological charges $|\ell| = \pm 3$.

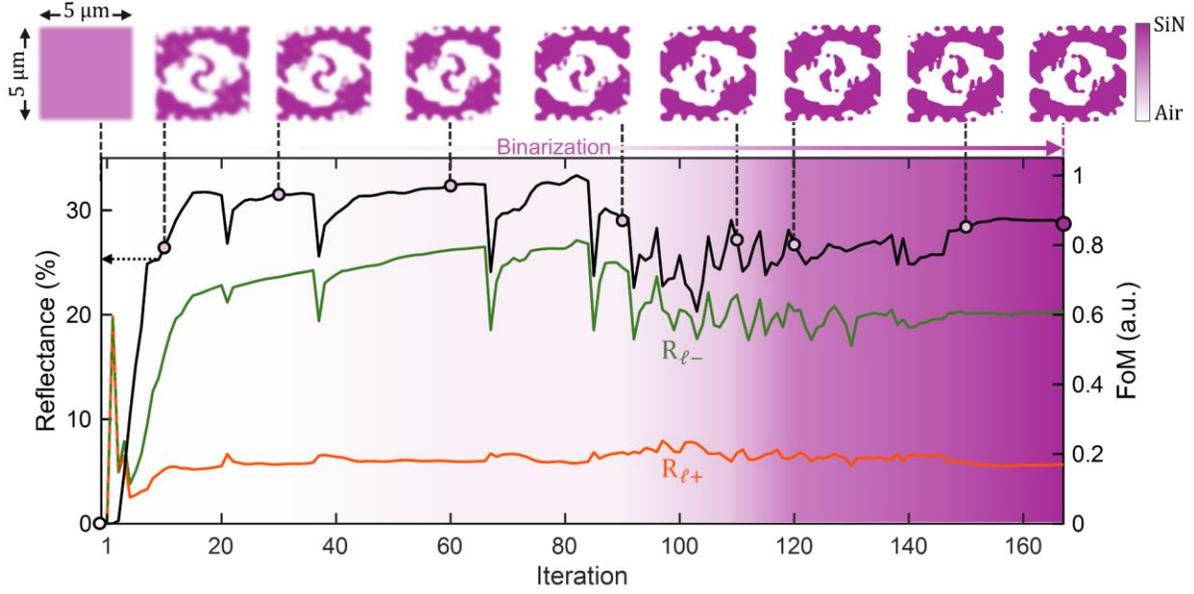

**Figure 3.** Iterative evolution of the figure of merit (FoM, black) alongside the reflectance for helical incidence beams with +3 and -3 topological charges ($R_{\ell+}$, orange; $R_{\ell-}$, green). The purple gradation in the background represents the degree of binarization across iterations. The oscillations in the curves indicate structural changes during the fabrication constraint process. The top subplots depict the transition from the initial grayscale structure to the final binarized configuration of SiN (purple) and Air (white), with each subplot connected by dashed lines to the corresponding iteration on the FoM curve. The design region is 5 μm × 5 μm, with a thickness of 800 nm. The minimum size of local features in the final design is restricted to 200 nm to maintain an aspect ratio of 4.

Each iterative optimization process begins with the forward simulations of the LG beam propagation with $|\pm\ell| = 3$, followed by the corresponding adjoint simulations to collect the material density gradient for HD maximization, as shown in **Fig. 2**. Then, the external gradient ascent algorithm updates the geometry of the chiral structure until the fabrication constraint factor [30] reaches a predefined threshold, and the algorithm no longer satisfies the Armijo sufficient condition [31]. **Figure. 3** illustrates the evolution of the FoM and the reflectance spectra as the number of iterations increases. In our case, after slightly more than 160 iterations, the change in FoM approaches nearly zero, and the geometry becomes fully binarized. The background purple gradation in **Fig. 3** intuitively represents the degree of binarization, illustrating how the refractive index profile transitions from a gray, continuous state to a fully binary, discrete state during the optimization process. The top subplots of **Fig. 3** visualize the transition from the initial 5 μm × 5 μm grayscale structure to the final binarized design of Si$_3$N$_4$ (purple) and Air (white). Each subplot is connected by dashed lines to its corresponding iteration number. $R_{\ell-}$ (green line) and $R_{\ell+}$ (orange line) represents the reflectance spectra for incident OAM beams with −3 and +3 topological charge modes, respectively. As the iteration process approaches the final stage, $R_{\ell-}$ exhibits approximately 20% reflectance, while only 6% reflectance is observed for $R_{\ell+}$, clearly indicating that our inverse design method successfully developed the HD freeform structure.

To evaluate the performance of the inverse-designed chiral structure in **Fig. 3**, we import the structure into a new simulation environment and place it on a glass substrate with substrate thickness H$_2$ = 600 nm as shown in **Fig. 4a**. The simulation domain incorporates the incident OAM beams with topological charges, $|\pm\ell|$, varying from 0 to 10. **Fig. 4b** and **4c** show the time-averaged electric field distributions for the incident OAM beams with $\ell = +3$ and $\ell = -3$, respectively. The series of images at various cross-sections visualize the light interaction above and within the chiral structure. Both the chiral structure and the field profile show the 2-fold rotational symmetry (C$_2$ symmetry), which we intentionally implemented in the optimization constraint. **Figure 4** shows the electric field intensity profile at various monitor locations, i.e., z = 0, 0.2, 0.4, 0.6, 0.8, and 1.6 μm, for the positive topological charge (**Fig. 4b**) and the negative topological charge (**Fig. 4c**). The near-field images differ based on the sign of the topological charge of the input beam, producing a dichroic response, as confirmed by the reflection intensity at z=1.6 μm. Interestingly, the field intensity profile of topological charge $\ell = -3$ at the reflection plane denoted with the blue outline in **Fig. 4a** presents the clear diagonal (y = x) line distribution, possibly due to the C$_2$ symmetric shape of the freeform structure. We conducted additional simulations using Lumerical FDTD Solutions software to ensure that the results are consistent and robust against potential slight variations in simulation settings across different platforms. This helically dichroic behavior is demonstrated in **Fig. 4d**, which shows the difference between positive and negative topological charges $|\pm\ell|$ from 0 to 10. MEEP simulation results are represented by orange and green lines with circular data points (subscript A), while Lumerical simulation results are represented by light green and pink lines with triangular data points (subscript B). Both results show the largest reflectance difference at $|\pm\ell| = 3$, which is in good agreement with the desired performance when optimizing the structure under the incident OAM mode $|\pm\ell| = 3$. **Figure. 4e** results display a peak HD at topological charge 3, where the reflectance difference is maximized, with the numerically highest HD signal of ~107% at 800 nm wavelength. Note that if topological charges $|\pm\ell|$ are not at 3, the HD responses exhibit ~75% at topological charges $|\pm\ell| = 4$ and still show ~30% when topological charges $|\pm\ell| = 7$. The performance suggests that the chiral structure exhibits broadband helically dichroic responses across various OAM topological charges. Both simulation tools show the same trend in helical dichroism as the topological charge varies. The consistency between these two independent numerical implementations strengthens the reliability of our computational predictions and supports the robustness of the designed chiral structure's performance.

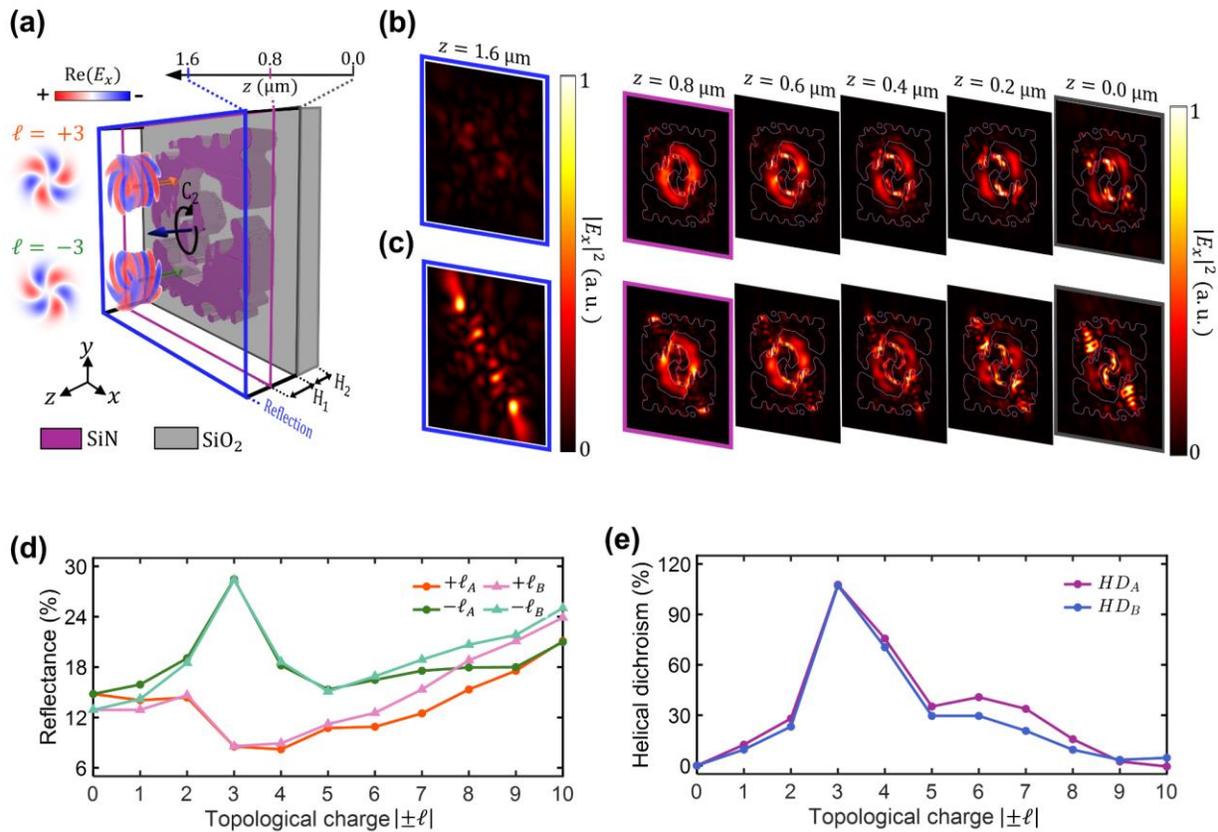

**Figure 4.** Numerical simulation of the optimized freeform chiral structure. **(a)** Schematic of the simulation domain showing the interaction between the incident OAM beam with topological charges $|\pm\ell| = 3$ and the chiral structure on a SiO$_2$ substrate. The field profiles on the left represent the real part of the electric field distribution of OAM beams. The beam diameter is 1.6 µm. Field monitors are placed along the z-axis, with the reflection monitor at z = 1.6 µm (blue outline). The chiral structure height (H1) and substrate thickness (H2) are 800 nm and 600 nm, respectively. **(b),**

**(c)** Time-averaged electric field intensity profiles for $\ell = +3$ and $\ell = -3$, measured at different z-positions. The significant contrast in the reflected fields (blue outline) between $\ell = +3$ and $\ell = -3$ inputs demonstrates a strong helical dichroism. **(d)** Simulated reflectance as a function of topological charge magnitude $|\pm\ell|$ 0 to 10 for positive and negative topological charges. Subscript A represents the MEEP, and B stands for Lumerical simulations. **(e)** Corresponding helical dichroism spectrum as a function of $|\pm\ell|$, showing maximum HD at $|\pm\ell| = 3$ for both MEEP and Lumerical simulations.

## Conclusion

In this report, we present the first-ever inverse-designed chiral structure to enhance the contrast in helical dichroism. The adjoint optimization method, known for its effectiveness in various design problems, was employed in this study. This approach resulted in a threefold increase in helical dichroism compared to conventional chiral structures, such as spiral arms. By leveraging adjoint optimization methods, we designed a chiral structure that exhibits a record-breaking helical dichroism response of ~107% for topological charges $\ell = \pm 3$ at 800 nm. The HD value can be further improved by relaxing constraints used in our simulation, such as the minimum feature size, height, and material. Inverse design is especially effective when an intuitive design strategy or governing physics principles are lacking, as in the case of chirality design. This study expands the promising applications of inverse design to helical dichroism and opens new possibilities for practical applications ranging from sensitive chiral molecular detection and advanced spectroscopic techniques to innovative pharmaceutical development.

## Acknowledgments

This research was supported by the Basic Science Research Program through the National Research Foundation of Korea(NRF), funded by the Ministry of Education (RS-2024-00405947), (RS-2024-00338048), and (RS-2024-00414119). It was also supported by the Global Research Support Program in the Digital Field (RS-2024-00412644) under the supervision of the Institute of Information and Communications Technology Planning & Evaluation (IITP), and by the Artificial Intelligence Graduate School Program (RS-2020-II201373, Hanyang University), also supervised by the IITP. Additionally, this research was supported by the Artificial Intelligence Semiconductor Support Program (RS-2023-00253914), funded by the IITP, and by the Korea government (MSIT) grant (RS-2023-00261368). This work received support from the Culture,

Sports, and Tourism R&D Program through a grant from the Korea Creative Content Agency, funded by the Ministry of Culture, Sports and Tourism (RS-2024-00332210).

## References

1. Cho, N.H., et al., *Bioinspired chiral inorganic nanomaterials.* Nature Reviews Bioengineering, 2023. **1**(2): p. 88-106.
2. Evans, A.M., *Comparative pharmacology of S (+)-ibuprofen and (RS)-ibuprofen.* Clinical rheumatology, 2001. **20**: p. 9-14.
3. Hamidi, S. and A. Jouyban, *Pre-concentration approaches combined with capillary electrophoresis in bioanalysis of chiral cardiovascular drugs.* Pharmaceutical Sciences, 2015. **21**(4): p. 229-243.
4. Oh, S.S. and O. Hess, *Chiral metamaterials: enhancement and control of optical activity and circular dichroism.* Nano Convergence, 2015. **2**: p. 1-14.
5. Zhu, A.Y., et al., *Giant intrinsic chiro-optical activity in planar dielectric nanostructures.* Light: Science & Applications, 2018. **7**(2): p. 17158-17158.
6. Yu, C.-L., et al., *High circular polarized nanolaser with chiral gammadion metal cavity.* Scientific reports, 2020. **10**(1): p. 7880.
7. Woody, R.W., *[4] Circular dichroism.* Methods in enzymology, 1995. **246**: p. 34-71.
8. Berova, N., K. Nakanishi, and R.W. Woody, *Circular dichroism: principles and applications*. 2000: John Wiley & Sons.
9. Lininger, A., et al., *Chirality in light–matter interaction.* Advanced Materials, 2023. **35**(34): p. 2107325.
10. Warning, L.A., et al., *Nanophotonic approaches for chirality sensing.* ACS nano, 2021. **15**(10): p. 15538-15566.
11. Duan, X., S. Yue, and N. Liu, *Understanding complex chiral plasmonics.* Nanoscale, 2015. **7**(41): p. 17237-17243.
12. Kwon, D.-H., P.L. Werner, and D.H. Werner, *Optical planar chiral metamaterial designs for strong circular dichroism and polarization rotation.* Optics express, 2008. **16**(16): p. 11802-11807.
13. Cao, T., et al., *Strongly tunable circular dichroism in gammadion chiral phase-change metamaterials.* Optics express, 2013. **21**(23): p. 27841-27851.
14. Bian, W., et al., *Sandwich-type planar chiral metamaterials for exploring circular dichroism.* Plasmonics, 2024. **19**(1): p. 389-394.
15. Shen, Y., et al., *Optical vortices 30 years on: OAM manipulation from topological charge to multiple singularities.* Light: Science & Applications, 2019. **8**(1): p. 90.
16. Babiker, M., et al., *Orbital angular momentum exchange in the interaction of twisted light with molecules.* Physical review letters, 2002. **89**(14): p. 143601.
17. Allen, L., et al., *Orbital angular momentum of light and the transformation of Laguerre-Gaussian laser modes.* Physical review A, 1992. **45**(11): p. 8185.
18. Andrews, D.L. and M. Babiker, *The angular momentum of light*. 2012: Cambridge University Press.
19. Mun, J., et al., *Electromagnetic chirality: from fundamentals to nontraditional chiroptical phenomena.* Light: Science & Applications, 2020. **9**(1): p. 139.

20. Wu, T., R. Wang, and X. Zhang, *Plasmon-induced strong interaction between chiral molecules and orbital angular momentum of light.* Scientific Reports, 2015. **5**(1): p. 18003.
21. Ni, J., et al., *Giant helical dichroism of single chiral nanostructures with photonic orbital angular momentum.* ACS nano, 2021. **15**(2): p. 2893-2900.
22. Dai, N., et al., *Robust Helical Dichroism on Microadditively manufactured copper helices via photonic orbital angular momentum.* ACS nano, 2023. **17**(2): p. 1541-1549.
23. Lim, Y.C., et al., *Strong Chiral Response of Chiral Plasmonic Nanoparticles to Photonic Orbital Angular Momentum.* Advanced Optical Materials: p. 2402268.
24. Lalau-Keraly, C.M., et al., *Adjoint shape optimization applied to electromagnetic design.* Optics express, 2013. **21**(18): p. 21693-21701.
25. Miller, O.D., *Photonic design: From fundamental solar cell physics to computational inverse design*. 2012: University of California, Berkeley.
26. White, A.D., et al., *Inverse design of optical vortex beam emitters.* ACS Photonics, 2022. **10**(4): p. 803-807.
27. Molesky, S., et al., *Inverse design in nanophotonics.* Nature Photonics, 2018. **12**(11): p. 659-670.
28. Bae, M., et al., *Inverse design and optical vortex manipulation for thin-film absorption enhancement.* Nanophotonics, 2023. **12**(22): p. 4239-4254.
29. Oskooi, A.F., et al., *MEEP: A flexible free-software package for electromagnetic simulations by the FDTD method.* Computer Physics Communications, 2010. **181**(3): p. 687-702.
30. Longman, A. and R. Fedosejevs, *Mode conversion efficiency to Laguerre-Gaussian OAM modes using spiral phase optics.* Optics Express, 2017. **25**(15): p. 17382-17392.
30. Hammond, Alec M., et al. "High-performance hybrid time/frequency-domain topology optimization for large-scale photonics inverse design. Optics Express, 2022. 30(3): p. 4467-4491.
31. Asl, Azam, and Michael L. Overton. "Analysis of the gradient method with an Armijo–Wolfe line search on a class of non-smooth convex functions. *Optimization methods and software,* 2020. 35(2): p. 223-242.